\documentclass[reprint,amsmath, twocolumn, amssymb, aps, notitlepage, superscriptaddress, pra]{revtex4-2}

\usepackage{amsmath,amsfonts}    
\usepackage{graphicx}   
\usepackage{verbatim}   
\usepackage{color,xcolor,soul}      
\usepackage{subfigure}  
\usepackage[colorlinks=true, citecolor=blue, urlcolor=blue]{hyperref}
\usepackage{textcomp}
\usepackage{bm}
\usepackage{bbm}
\usepackage{braket}
\usepackage[normalem]{ulem}
\usepackage{natbib}
\sethlcolor{yellow}

\newcommand{\be}{\begin{equation}}
\newcommand{\ee}{\end{equation}}
\newcommand{\ba}{\begin{array}}
\newcommand{\ea}{\end{array}}
\newcommand{\bqa}{\begin{eqnarray}}
\newcommand{\eqa}{\end{eqnarray}}

\graphicspath{{figs/}}

\definecolor{gr}{RGB}{225,225,225}

\renewcommand{\selectlanguage}[1]{}

\begin{document}

\title{Quantum information spreading in inhomogeneous spin ensembles}

\author{Rahul Gupta}
\affiliation{Department of Physics, Indian Institute of Technology Bombay, Powai, Mumbai 400076, India}
\affiliation{Physics Department, Blackett Laboratory, Imperial College London, Prince Consort Road, SW7 2AZ, United Kingdom}

\author{Florian Mintert}
\affiliation{Physics Department, Blackett Laboratory, Imperial College London, Prince Consort Road, SW7 2AZ, United Kingdom}

\author{Himadri Shekhar Dhar}
\affiliation{Department of Physics, Indian Institute of Technology Bombay, Powai, Mumbai 400076, India}
\affiliation{Centre of Excellence in Quantum Information, Computation, Science and Technology, Indian Institute of Technology Bombay, Mumbai 400076, India}

\date{\today}

\begin{abstract}
We present a Krylov space based theoretical framework for modeling inhomogeneous spin ensembles with arbitrary distributions of spin frequencies and couplings. 
The framework is then used to
{asymptotically large spin ensemble}.
{In the single-excitation subspace, the Krylov construction allows for to derive exact expressions}
for
the Lieb-Robinson velocity and {quantum speed limit}, 
and
figure of merit such as Krylov complexity.
{Our work reveals a strong dependence of the speed of information flow on the statistical distribution of resonance frequencies in the spin ensemble with immediate implications for the design of components for quantum technologies, realized 
for example with}  
nitrogen vacancy centers, 
nuclear spins or ultracold atoms.   
\end{abstract}


\maketitle 

\section{Introduction\label{sec:intro}}
Understanding how quantum information spreads in many-body systems~\cite{Eisert2015,Iyoda2018,Abanin2019} is a central problem in contemporary quantum science, lying at the intersection of quantum information theory~\cite{Luitz2017,Nahum2018}, condensed matter physics~\cite{Swingle2017}, and quantum optics~\cite{Haroche2006}. 
The dynamics of information propagation under unitary evolution governs fundamental processes such as thermalization~\cite{Deutsch2018}, scrambling~\cite{Swingle2017}, entanglement growth~\cite{Ho2017}, and state transfer~\cite{Matsukevich2004}, and it also sets practical limits for quantum technologies such as quantum memories~\cite{Xu2014,Paulson2022}, quantum communication networks~\cite{Murphy2010}, and 
hybrid architectures~\cite{Kurizki2015,Clerk2020}.
In recent years, significant progress has been made in characterizing information spreading using tools such as Lieb–Robinson bounds (LRBs)~\cite{Lieb1972,Hastings2004},
{quantum speed limits (QSL)~\cite{Mandelstam1991,Margolus1998}, out-of-time-order correlators~\cite{Swingle2018}, operator growth~\cite{Parker2019} and, more recently, Krylov complexity~\cite{Rabinovici2022}.} 
These approaches have provided new geometric~\cite{Pires2016} and dynamical perspectives on how quantum states evolve in Hilbert space and how fast information can propagate under physical constraints.

Among these developments, the Krylov-space formulation of quantum dynamics has emerged as a particularly powerful framework~\cite{Takahashi2025,Nizami2023}. By mapping unitary evolution onto an effective one-dimensional tight-binding problem in Krylov space, 
one can describe the growth of operators or states in terms of Krylov basis states and Krylov coefficients~\cite{Rabinovici2021,Nandy2025}. This representation has enabled quantitative studies of information spreading, wavepacket propagation in Krylov space, and measures such as Krylov complexity, which captures how rapidly a quantum state explores larger regions of Hilbert space~\cite{Caputa2022,Choudhary2026}. 
On the other hand, quantum speed limits, such as the Mandelstam–Tamm (MT) and Margolus–Levitin (ML) bounds~\cite{Mandelstam1991,Margolus1998}, have been extensively studied as fundamental constraints on the minimal time required for a quantum system to evolve between two states with a given fidelity. While these two lines of research have largely progressed independently, their synthesis offers a promising route to obtain fine-grained, state-resolved bounds on quantum information dynamics~\cite{Bhattacharya2024,Gill2025}.

Spin ensembles coupled to a central mode, such as a cavity photon or a qubit, form an important class of {hybrid quantum systems, where questions related to spread of information}
are both theoretically rich and experimentally relevant. Such models underpin a wide range of platforms, including {cavity quantum electrodynamics} (QED)~\cite{Haroche2006,Mivehvar2021}, circuit QED~\cite{Kubo2011,Blais2021}, 
and quantum magnonics~\cite{LachanceQuirion2019,Skogvoll2021}. 
However, realistic spin ensembles are inherently inhomogeneous -- individual spins typically have different transition frequencies and coupling strengths due to disorder, fabrication imperfections~\cite{Levchenko2015}, or spatial variation~\cite{Rosenzweig2018}. 
This inhomogeneity profoundly affects coherence, transport, and information storage, often leading to dephasing and loss of retrievability~\cite{Dorovitski2008}. While homogeneous or weakly disordered models admit elegant analytical treatments~\cite{Keeling2009,Zeb2022,Sharma2025} using Holstein-Primakoff 
bosonization~\cite{Diniz2011,Wesenberg2011}, 
inhomogeneous spin ensembles remain challenging, especially when one seeks exact results rather than mean-field~\cite{Wesenberg2011,Kurucz2011} or numerical approximations~\cite{Ansel2018}.

Existing studies of information spreading and quantum speed limits in spin ensembles have therefore faced several limitations. {Firstly}, most analytical results rely on assumptions of uniform couplings or identical spin frequencies,
{largely ignoring the effect of physical imperfections or disorder in the system \cite{Chase2008,Sharma2025_p}.}
{Secondly}, even when inhomogeneity is included, analyses are often restricted to mean-field analysis or restricted to a small number of spins \cite{Iemini2024}.
{Finally}, although Krylov methods have been applied to operator growth and chaos, analytical construction of the Krylov space are rare in the context of spin ensembles and hybrid quantum systems, and are typically limited to special models~\cite{Takahashi2025}. 

In this work, we address these challenges by developing a 
Krylov-space description of quantum information spreading in {single-excitation manifold of}
{an} inhomogeneous spin ensemble. 
{The theoretical framework captures the genuinely quantum dynamics of a single shared excitation, where} 
both the spin–photon
couplings
and the spin transition frequencies are allowed to be completely inhomogeneous.
The exact Krylov basis states and Hamiltonian {are constructed to provide an} 
analytically tractable representation of the dynamics. 
{This} allows {for}
{the derivation of
analytical expressions for the figures of merit such as Lieb-Robinson velocity, quantum speed limit (QSL) and Krylov complexity, which captures the flow of information in the Krylov space.}
{The information dynamics is explored for different inhomogeneous spin ensembles, ranging from uniform to $q$-Gaussian distribution of spin frequencies, and the spread or localization of information is captured in each case.}
The findings are directly relevant for the design and optimization of quantum memories and interfaces, where controlling information spreading and minimizing loss due to disorder are critical. More broadly, this work establishes a powerful framework for studying quantum dynamics in disordered systems, highlighting how the structure of underlying distributions governs information propagation at a fundamental level.

The
paper is arranged as follows. {Section~\ref{sec:setup} describes the construction of Krylov space. The figures of merit to study the spread of information and their theoretical bounds are discussed in Sec.~\ref{sec:info_spread}.
The calculation of the exact Krylov states and the single excitation dynamics
for centrally symmetric, $q$-Gaussian, and uniform distributions are shown in Sec.~\ref{sec:Krylov_coeff}. {The physical systems corresponding to the different spin distributions
are} discussed in Sec.~\ref{sec:physical}, with conclusion and outlook in  
Sec.~\ref{sec:conc}.}

\section{Modelling Krylov Subspace Hamiltonian \label{sec:setup}}


The dynamics of an inhomogeneously broadened spin ensemble coupled to a single-mode optical cavity can be described by the Tavis–Cummings model (TCM)~\cite{Tavis1968}, 
whose Hamiltonian is written as $\mathcal{H} = \mathcal{H}_0 + \mathcal{H}_I$, where, the free (non-interacting) contribution is given by
\begin{equation}
\mathcal{H}_0 = \omega_c a^\dagger a + \sum_{j=1}^{N}
{\omega_j\sigma_j^+\sigma_j^-},
\label{eq:iTCMfree}
\end{equation}
while the interaction between the cavity field and the spin ensemble takes the form
\begin{equation}
\mathcal{H}_I =
\sum_{j=1}^{N} g_j \left(a^\dagger \sigma_j^- + a \sigma_j^+\right).
\label{eq:iTCMint}
\end{equation}
{Here} $a$ and $a^\dagger$ denote the annihilation and creation operators of the cavity mode, and $\sigma_j^\pm$ and $\sigma_j^z$ represent the raising (lowering) and Pauli-$z$ operators for the $j^{th}$ spin, respectively. {The individual spin-cavity coupling is given by $g_j$. The TCM Hamiltonian conserves the total excitations in the system i.e. 
{$N_{\rm ex} = n + n_e$,
where $n$ and $n_e$ are the number of photons and excited spins, respectively}. 

In practice, while the cavity frequency $\omega_c$ is typically well characterized, the large number of spins makes a precise determination of the individual spin transition frequencies $\omega_j$ and coupling strengths $g_j$ in an inhomogeneous ensemble infeasible. 
{For a typical spin ensemble, the Hilbert space grows with the number of spins even for low-excitation subspaces and when combined with the uncertainty in these microscopic parameters, renders direct numerical simulations prohibitively difficult.}
These challenges can be addressed by constructing an effective description in terms of a Krylov subspace, {where the relevant dynamics can be accurately captured in a significantly reduced basis.}


{In the single-excitation subspace of the Hilbert space,} where the total number of excitations satisfies $N_{\mathrm{ex}} = 1$, 
the Hamiltonian $\mathcal{H}$ in Eqs.~\eqref{eq:iTCMfree}-\eqref{eq:iTCMint} can be written using as
\begin{equation}
\mathcal{H}_r=\omega_c\vert 0\rangle\langle0\vert + \sum_{j=1}^N\left[ \omega_j \vert j\rangle\langle j\vert +  g_j \left(\vert0\rangle\langle j\vert+\vert j\rangle\langle 0\vert\right)\right],
\label{eq:H_single_ex}
\end{equation}
where
$\vert 0\rangle=\vert 1\rangle_c\otimes\vert 0 {\rangle_s^{\otimes N}}$
is the joint state with single excitation in cavity mode and all $N$ spins in ground state, while ${\vert j\rangle=\vert0\rangle_c\otimes\vert 0\rangle^{\otimes j-1}_s\otimes\vert1\rangle_s\otimes\vert0\rangle^{\otimes N-j}_s}$ is the joint state where {only} the $j^{th}$ spin with frequency $\omega_j$ is excited, while the cavity is in vacuum.

With the restriction to the single-excitation subspace, the dimension of the Hilbert space grows only linearly in the number of spins, but with typical numbers in the range of $10^{12}-10^{16}$ even such a moderate scaling results in untractable system sizes.
More importantly, the detailed system dynamics will depend on the actual value of each individual resonance frequency $\omega_j$, and a control protocol that assumes knowledge of all these frequencies does not provide a practical way forward.

A representation in a Krylov basis addresses both of these issues.
It does provide a basis that admits a truncation to a low-dimensional subspace of the Hilbert space in which the most relevant dynamics takes place, and the matrix elements of the Hamiltonian in this basis do not depended on the individual spin frequencies $\omega_j$, but only on properties of their statistical distribution, such as their average, variance and potentially higher-order statistical moments.
%
The underlying idea is to start with a set of states {described by the images of an initial state $|\psi\rangle$ under increasing powers of the Hamiltonian $\mathcal{H}_r$. The Krylov space is then spanned by}
\be
{\{|\psi\rangle,\mathcal{H}_r|\psi\rangle,\mathcal{H}_r^2|\psi\rangle,\dots,\mathcal{H}_r^{M-1}|\psi\rangle \}.}
\label{eq:Krylov_span}
\ee
{The initial $\ket{\psi}$ can be an arbitrary state in the single-excitation space ($N_{\rm{ex}}=1$)
and} $M \ll N$, which leads to a significant reduction in the effective Hilbert space.

To systematically construct an orthonormalized Krylov basis $\{|\phi_n\rangle\}_{n=0}^{M-1}$ from the span in Eq.~\eqref{eq:Krylov_span}, a useful tool is the Lanczos algorithm~\cite{Lanczos1950}, given by
\begin{align}
&\vert\tilde{\phi}_{n+1}\rangle=\mathcal{H}_r\vert\phi_{n}\rangle - \alpha_n\vert\phi_n\rangle-\beta_n\vert\phi_{n-1}\rangle,
\label{eq:krylov}
\end{align}
where $\vert\tilde{\phi}_{n+1}\rangle$ is orthogonal to $\vert\phi_{n}\rangle$ and is normalized by $\vert\phi_{n+1}\rangle=\vert\tilde{\phi}_{n+1}\rangle/\beta_{n+1}$, with $\beta^2_n=\langle\tilde{\phi}_n\vert\tilde{\phi_n}\rangle$. Now, the diagonal elements of 
the Hamiltonian $\mathcal{H}_r$ in the orthonormal Krylov basis
{are} given by the coefficients
\begin{equation}
\alpha_n=\langle\phi_n\vert \mathcal{H}_r\vert \phi_n\rangle. \label{eq:diag}
\end{equation}
{The off-diagonal elements can be obtained by rearranging} Eq.~\eqref{eq:krylov} such that
\begin{eqnarray}
&\mathcal{H}_r\vert\phi_{n}\rangle = \beta_{n+1}\vert{\phi}_{n+1}\rangle + \alpha_n\vert\phi_n\rangle + \beta_n\vert\phi_{n-1}\rangle,~\textrm{s.t.}&\\
&\langle\phi_{n+1}\vert\mathcal{H}_r\vert\phi_{n}\rangle = \beta_{n+1} ~~\textrm{and}~~ \langle\phi_{n-1} \vert\mathcal{H}_r\vert\phi_{n}\rangle = \beta_{n}. \label{eq:offdiag}&
\end{eqnarray}
All other off-diagonal terms by definition vanish.
Now,
the Hamiltonian $\mathcal{H}_r$ in the Krylov basis
{is} a tridiagonal matrix, {which} can be written as 
\begin{align}
    \mathcal{H}_k=\begin{pmatrix}
        \alpha_0 & \beta_1 & 0 & 0  & ... & 0 & 0 & 0\\
        \beta_1 & \alpha_1 & \beta_2 & 0  & ... & 0 & 0& 0\\
        0 & \beta_2 & \alpha_2 & \beta_3  & ... & 0 & 0& 0\\
        . & . & . & . & . & . & . & .\\ 
        . & . & . & . & . & . & . & .\\ 
        0 & 0 & 0 & 0 & ... & \beta_{M-1}& \alpha_{M-1} & \beta_{M}\\
        0 & 0 & 0 & 0 & ... & 0 & \beta_{M} & \alpha_{M}
    \end{pmatrix}.
\label{eq:KrylovHam}
\end{align}

{The first few Krylov states in the joint single excitation basis can be evaluated}
using the Hamiltonian in Eq.~\eqref{eq:H_single_ex}.
The first normalized Krylov state is $\vert\phi_0\rangle=\vert 0\rangle$, which represents the spin ground state with a single photon in the cavity. Using Eq.~\eqref{eq:krylov}, the unnormalized state, orthogonal to $\vert\phi_0\rangle$, is $\vert\tilde{\phi}_1\rangle=\sum_{j}g_j\vert j\rangle$, which is the collective bright state.
Now, from Eqs.~\eqref{eq:diag}-\eqref{eq:offdiag}, the first terms are $\alpha_0=\omega_c$, $\beta_1=\sqrt{\sum_j g_j^2}=g_{\rm{eff}}$ and 
$\alpha_1=\dfrac{\sum_{j}g_j^2\omega_j}{\sum_{j}g_j^2}=\bar{\omega}$. 
The following {Krylov} states can  similarly be derived using {the recursion in Eq.~\eqref{eq:krylov}:}
\begin{eqnarray}
    \ket{\tilde{\phi}_2}=\beta_2\ket{\phi_2}&=&(\mathcal{H}_r-\alpha_1)\ket{\phi_1}-\beta_1\ket{\phi_0}, \textrm{where} \label{eq:phi_2}\\
    \mathcal{H}_r\ket{\phi_1} &=& \sum_j(g_j\omega_j\ket{j} + g^2_j\ket{0})/\beta_1 \label{eq:Hr_phi1}.
\end{eqnarray}
{The last terms in Eqs.~\eqref{eq:phi_2} and \eqref{eq:Hr_phi1} are both $\beta_1\ket{0}$ and cancel, and therefore}
\begin{eqnarray}
    &\vert\tilde{\phi_2}\rangle=\sum_j\dfrac{g_j}{\beta_1}(\omega_j-\bar{\omega})\vert j\rangle=\sum_j\dfrac{g_j}{\beta_1}\Delta_j\vert j\rangle,&\\
    &\beta_2^2=\sum_j\dfrac{g_j^2\Delta_j^2}{\beta_1^2},~~\alpha_2=\sum_j\dfrac{g_j^2\Delta_j^2\omega_j}{\beta_1^2\beta_2^2}.&
\end{eqnarray}
Typically, spin ensembles with finite, but large $N$, have centrally symmetric spin frequency distributions, {with an independent distribution of coupling strengths~\cite{Putz2014,Krimer2014}.}
As such, for joint distributions, symmetric about mean spin frequency $\bar{\omega}$, the moments $\textbf{E}(\Delta^n)=\sum_jg_j^{2}\Delta_j^n$, where  $\Delta_j^n = (\omega_j - \bar{\omega})^n$, vanish for odd $n$.
%
%
Defining the spin frequency variance as $\sigma_\omega^2=\textbf{E}( \Delta^2)/\sum_j g_j^2=\sum_j g^2_j \Delta_j^2/\sum_j g_j^2$, 
\begin{align}
    \beta^2_2&=\frac{\sum_j g_j^2\Delta_j^2}{\sum_j g_j^2}=\sum_j \frac{g_j^2\Delta_j^2}{ \beta_1^2}=\sigma_\omega^2 ,\\
    \alpha_2&=\sum_j\frac{g_j^2\Delta_j^2\omega_j}{\beta_1^2\beta_2^2}
    =\frac{\textbf{E}(\Delta^3)}{\beta_2^2} + \bar{\omega}\frac{\textbf{E}(\Delta^2)}{\beta_2^2}=\bar{\omega},\label{eq:alpha20}
\end{align}
where, in the second term,
$\omega = \bar{\omega}+\Delta$.
Again using Eqs.~\eqref{eq:diag}-\eqref{eq:offdiag},
\begin{align}
    \beta_3^2&=\sum_j\frac{g_j^2(\Delta_j^2-\beta_2^2)^2}{\beta_2^2\beta_1^2}=\frac{\textbf{E}((\Delta^2-\beta_2^2)^2)}{\beta_2^2},\\
    \alpha_3&=\sum_j\frac{g_j^2\omega_j(\Delta_j^2-\beta_2^2)^2}{\beta_3^2\beta_2^2\beta_1^2}
    =\bar{\omega}.
\end{align}
{As such, for a centrally symmetric spin distribution $\alpha_n=\bar{\omega}~\forall~n\geq1$, {with 
$\omega_c=\alpha_0=\bar{\omega}$ for a resonant cavity.}

{Importantly, all the higher Krylov states and associated coefficients depend explicitly on the statistical properties of the inhomogeneous spin frequency and spin-photon coupling distribution rather than the intractable individual frequencies $\omega_j$ and coupling $g_j$.}
The key problem then {simply} reduces to analytically or numerically estimating $\beta_n$, as done for different spin ensemble distributions in Sec.~\ref{sec:Krylov_coeff}.




\section{Characterizing Spread of Information\label{sec:info_spread}}

The matrix of the Hamiltonian $\mathcal{H}_k$ in Eq.~\eqref{eq:KrylovHam} can be written in an operator form as
\begin{align}
\mathcal{H}_k &=\sum_{i,j=0}^{M-1} H_{i,j} = \bar{\omega}~\mathbb{I} + \sum_{i=0}^{M-1} \beta_{i+1}X_{i,i+1},
\label{eq:spin_local_Ham}
\end{align}
where $X_{i,i+1}=\vert\phi_{i}\rangle\langle\phi_{i+1}\vert + \vert\phi_{i+1}\rangle\langle\phi_{i}\vert$,
is an operator that swaps the states:  
$\vert\phi_i\rangle\rightleftharpoons\vert\phi_{i+1}\rangle$, {and the diagonal $\alpha_n$ terms are 
taken to be $\bar{\omega}$. As such, the dynamics of the system is governed by the second term {that facilitates exchange of excitations between neighboring Krylov states.}


\begin{figure*}[htb]
    \centering
    \includegraphics[width=\linewidth]{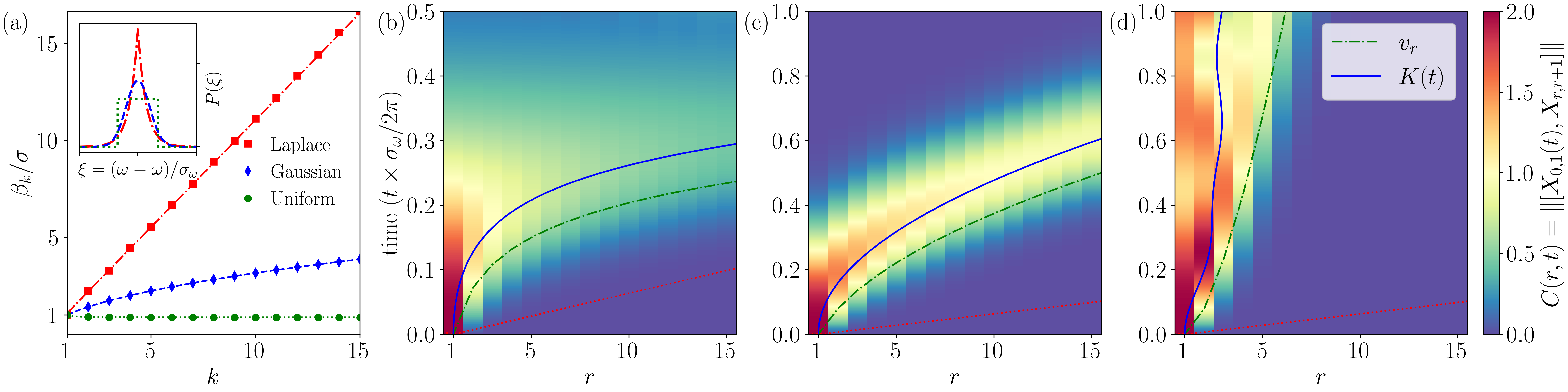}
    \caption{{Quantum information spreading in Krylov Space:} (a) Krylov coefficients $\beta_k$ for three frequency distributions calculated using Eq.~\eqref{eq:hankel_formulae}. The inset shows spin frequency distributions centered around mean frequency $\bar{\omega}$. (b,c,d) shows time evolution of local correlation function $C(r,t)=\left\|\left[X_{0,1}(t),X_{r,r+1}(t)\right]\right\|$ across Krylov space index separation $r$ for Laplace, Gaussian, and uniform distributions respectively. White dashed line shows the Krylov complexity $K(t)$ obtained from Eq.~\eqref{eq:krylov_complexity}, green dot-dashed line shows the local Lieb-Robinson velocities while red dotted line shows the global Lieb-Robinson bound for a Krylov subspace of fixed size $M=128$.}
    \label{fig:bounds}
\end{figure*}
{Two} key parameters used to study the spread of information are the Lieb–Robinson velocity~\cite{Lieb1972} and the Krylov complexity~\cite{Pires2016}.}
{The Lieb–Robinson velocity captures the local spread of correlations and information between the Krylov states $|\phi_i\rangle$ and $|\phi_{i+r}\rangle$, and is calculated by evaluating the correlator}
\begin{eqnarray}
C(r,t)&=&\|\left[X_{i,i+1}(t),X_{i+r,i+r+1}(0) \right]\|\nonumber\\
&=&a\exp{(-r+2Jt)},
\end{eqnarray}
{where $\|\cdot\|$ is the operator norm, $a$ is a constant prefactor, and $J =\|H_{i,i+1}\|$ is the uni-directional velocity with which the correlator $C(r,t)$ spreads between two Krylov states. The local  
velocity is given by the nearest-site norm in Eq.~\eqref{eq:spin_local_Ham} i.e., $v_i=2\|H_{i,i+1}\|=2|\beta_i|$. The maximum velocity corresponding to $J_{\rm{max}} = \rm{max}(|\beta_i|)$ gives the important Lieb-Robinson bound (LRB), which caps the rate at which information can spread in the system.}


{The other} figure of merit
is the Krylov complexity,
which is defined as
\begin{align}
K(t)&=\sum_n n \vert\langle\psi(t)\vert\phi_n\rangle\vert^2=\sum_{n}n|c_n(t)|^2.
\label{eq:krylov_complexity}
\end{align}
The Krylov complexity gives
the average Krylov state index or number that
{is}
populated during the time evolution, {thus estimating the spread of coefficients or delocalization of the state in the Krylov space.}
%
As such, Krylov complexity provides a different perspective from the 
Lieb-Robinson velocity,
which provides the
rate at which information spreads in the Krylov subspace.
{From an operational perspective, Krylov complexity is an essential tool to estimate the number of Krylov states necessary to simulate the dynamics of the system, while Lieb-Robinson velocity is an useful parameter in designing optimal control strategies.}
The variations of $K,v_i,$ and the LRB are shown {in Fig.~\ref{fig:bounds}} for different coefficients $\beta_k$ governed by the Laplace, Gaussian, and uniform distributions. While the LRB provides a light-cone-type global bound, $v_i$ characterizes the rate of spreading at each site, and $K$ denotes the spreading of the mean excitation in the Krylov subspace}. Thus, these quantities together provide a complete characterization of the 
{flow} of quantum information in spin ensembles.

{The speed of propagation of excitations} 
from Krylov state $\ket{\phi_i}$ to $\ket{\phi_j}$ {can also be studied in terms of fidelity $F_{ij}=|\langle\phi_i|\psi(t)\rangle|^2$, where for a complete transfer $\ket{\psi(t)}=\ket{\phi_j}$, the quantum speed limit is
given by the Mandelstam-Tamm (MT) bound~\cite{Mandelstam1991,Deffner2017}.} {The 
optimal evolution time corresponding to the bound is
\begin{align}
\tau_{ij}&=\frac{\arccos{\left(\sqrt{F_{ij}}\right)}}{\delta_{i} \mathcal{H}},~\textrm{where}\\
\delta_{i}\mathcal{H}&=\sqrt{\langle\phi_i\vert \mathcal{H}^2\ket{\phi_i} - \langle\phi_i\vert \mathcal{H}\ket{\phi_i}^2}.
\end{align}
The term $\delta_{i}\mathcal{H}$ can be calculated by using Eq.~\eqref{eq:spin_local_Ham}
\begin{equation}
\mathcal{H}\ket{\phi_i}=\sum_j \beta_{j+1}X_{j,j+1}\ket{\phi_i}
=\beta_i\ket{\phi_{i-1}}+\beta_{i+1}\ket{\phi_{i+1}}, 
\end{equation}
which gives $\langle\phi_i\vert H\ket{\phi_i}=0$ and $\langle\phi_i\vert H^2\ket{\phi_i}=\beta^2_i +\beta^2_{i+1}$. Therefore,
\begin{align}
\delta_{i}H&=\sqrt{\beta^2_i +\beta^2_{i+1}},~\tau_{i,j}=\frac{\arccos(\sqrt{F_{ij}})}{\sqrt{\beta^2_i +\beta^2_{i+1}}}.
\end{align}
{For bright mode as the initial state, the optimal time corresponding to the QSL is}
\begin{equation}
\tau_{1j}=\arccos\left(\sqrt{F_{1j}}\right)/\sqrt{g^2_{\rm{ens}}+\sigma_\omega^2},
\label{eq:QSL}
\end{equation}
which holds for any centrally symmetric frequency distribution with standard deviation $\sigma_\omega$. This suggests that orthogonal
state swapping cannot occur between the single photon and bright mode of spin ensemble in times which are smaller than $\tau_0=\pi/2\sqrt{g^2_{\rm{ens}}+\sigma_\omega^2}$. {As such,} evolution time $t$ is bounded by $t\geq\tau_0$. 
{Importantly, in the absence of a cavity and collective spin-photon coupling $g_{\rm{ens}}$,  
information in spin ensemble is effectively lost 
by $\tau_L=\pi/2\sigma_\omega$.} Thus, any control pulse {aiming at protecting information loss due to inhomogenity in the spin ensemble should 
operate within the temporal range $\tau_0\leq t_c\leq\tau_L$.} 


%

\section{Estimating the Krylov Coefficients\label{sec:Krylov_coeff}}

{An important set of quantities in studying the dynamics and the flow of information in an inhomogeneous spin ensemble
are the coefficients $\{\alpha_i,\beta_i\}$, which determine the Hamiltonian in Eq.~\eqref{eq:KrylovHam}.}
%
{These coefficients can be can readily calculated using the statistical properties of some physically relevant distributions pertaining to the spin ensemble.}

\subsection{General Frequency Distributions}
{For an inhomogeneous spin ensemble, where the spin frequencies follow a general probability distribution $P(\omega)$, its statistical properties can always be estimated using an orthogonal set of order $n$ polynomials $\{p_n(\omega)\}$~\cite{Konig2005}, where} 
\begin{align}
    \int_{-\infty}^\infty p_n(\omega)p_m(\omega)P(\omega)=
    \chi'_{n}\delta_{n,m},
    \label{chi_1}
\end{align}
%
%
{For a discrete distribution $P(\omega)=\sum_jg_j^2\delta(\omega-\omega_j)/g^2_{\rm{eff}}$, the orthogonalization is given by
\begin{align}
    \textbf{E}(p_n(\omega)p_m(\omega))&=\sum_{j=1}^N g_j^2 p_n(\omega_j)p_m(\omega_j )=\chi_{n}  \delta_{n,m},
    \label{chi_2}
\end{align}
where $\chi_{n} = \chi'_{n}g^2_{\rm{eff}}$ and $p_0(\omega)=1$.}
{These polynomials 
are directly related to the Lanczos algorithm for Krylov states in Eq.~\eqref{eq:krylov}, as from Favard's theorem~\cite{Favard1935}, $p_{n+1}(\omega)$  forms a three-term recursion relation unique for a given $P(\omega)$. This is given by}
\begin{align}
    p_{n+1}(\omega)=(\omega-\alpha_n)p_n(\omega) - \beta_n p_{n-1}(\omega),
\label{eq:three_term_rec}
\end{align}
where $\alpha_n,\beta_n$ are related to the moments of the distribution~\cite{Min2023,MAGNUS2008}
\begin{align}
    \alpha_n&=\frac{\textbf{E}\left[p_n(\omega)\omega p_n(\omega)\right]}{\textbf{E}\left[p_{n}(\omega)p_{n}(\omega)\right]},
    \beta_n=\frac{\textbf{E}\left[p_n(\omega)p_n(\omega)\right]}{\textbf{E}\left[p_{n-1}(\omega)p_{n-1}(\omega)\right]},
\label{eq:hankel_formulae}
\end{align}
with $\beta_n= {\chi_n}/{\chi_{n-1}}$. 
{For example, representing the Krylov states as $|\tilde{\phi}_n\rangle = \sum_jg_jp_n(\omega_j-\bar{\omega})\vert j\rangle$ in Eq.~\eqref{eq:krylov} results in the recursion relation in Eq.~\eqref{eq:three_term_rec}, providing a direct mapping between the polynomials and Krylov states and the coefficients $\{\alpha_i,\beta_i\}$.}

{For Gaussian distribution, the associated polynomials are the well-known Hermite polynomials $H_n(\omega_j-\bar{\omega})$, with $\chi_n' = \sigma_\omega^{2n} n!$~\cite{gautschi1996}. 
From the recursion in Eq.~\eqref{eq:three_term_rec}:}
\begin{align}
    \textbf{E}\left[H_n(\omega-\bar{\omega})\omega H_n(\omega-\bar{\omega})\right] 
    =\chi_{n}\bar{\omega},
\end{align}
{which leads to $\alpha_n = \bar{\omega}~\forall~n$. Moreover, $\beta_n={\chi'_n}/{\chi'_{n-1}} = \sigma_\omega \sqrt{n}$. With $\beta_0 = g_{\rm eff}$, the expressions for $\alpha_n$ and $\beta_n$, allow an exact analytical expression for the Hamiltonian $\mathcal{H}_k$ in the Krylov space. Interestingly, if the moments of the distribution are already known or can be computed easily, these terms can also be derived using the Henkel determinant formalism, as briefly described in Appendix~\ref{app:henkel}.}


\begin{figure*}[htb]
    \centering
    \includegraphics[width=\linewidth]{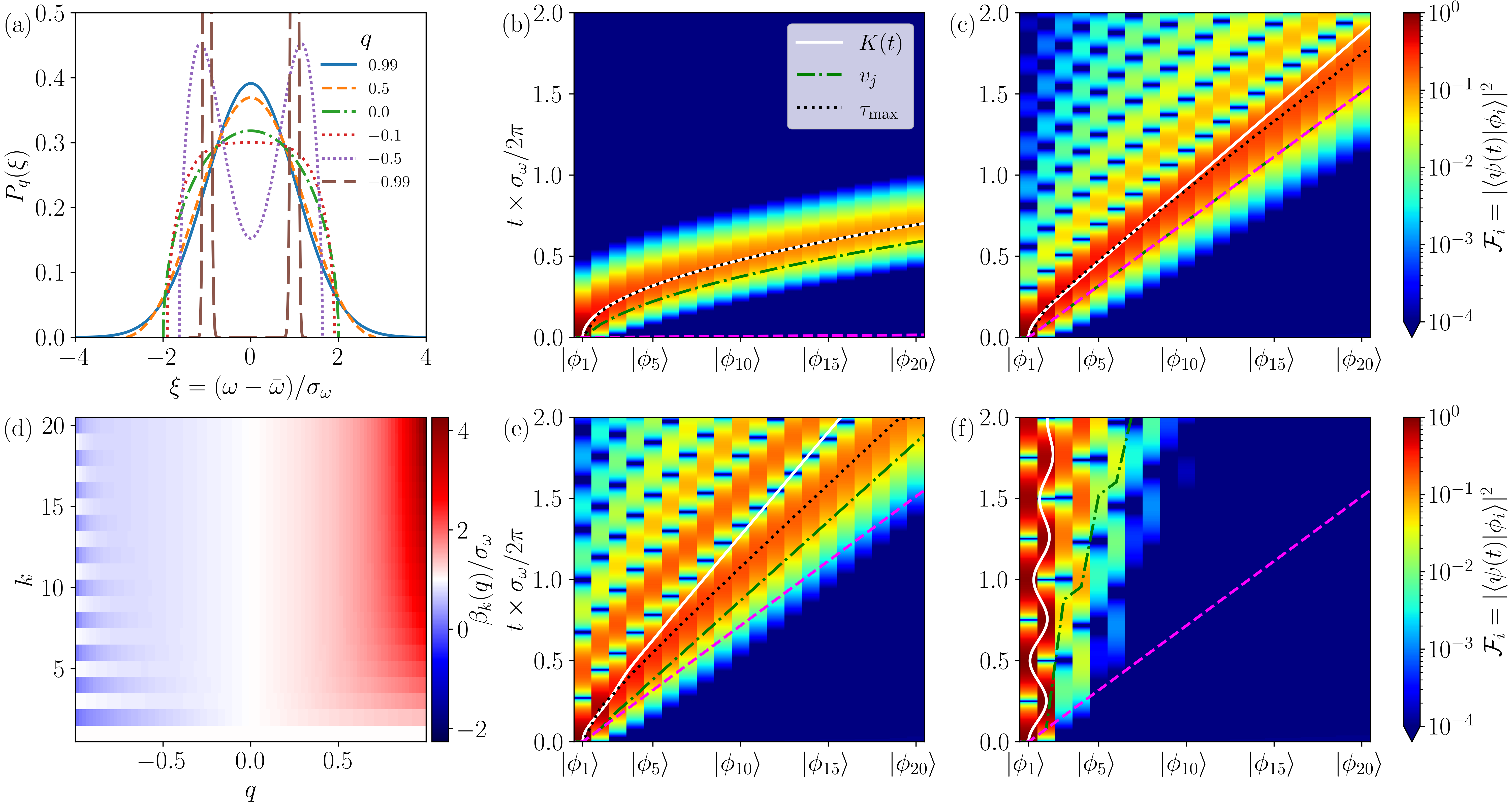}
    \caption{
    {Information dynamics in $q$-Gaussian spin ensembles:} (a) continuous $q$-Gaussian distributions for $-1\leq q\leq 1$ and (d) Krylov space coefficients $\{\beta_k(q)\}$ for different values of $q$. (b,c,e,f) shows time evolution of fidelity $\mathcal{F}_i=\vert\langle\psi(t)\vert\phi_i\rangle\vert^2$ of different Krylov states for $q=1,0,-0.5,-0.99$, respectively. The magenta dashed line shows Lieb-Robinson bound in the Krylov space, with white solid line showing the Krylov complexity $K(t)$ obtained from Eq.~\eqref{eq:krylov_complexity}. The green dot-dashed line shows the local Lieb-Robinson velocities $v_j$ and black dotted line shows optimal fidelity {time} for each state $\ket{\phi_j}$.}
    \label{fig:q-bounds_q_var}
\end{figure*}



\subsection{$q$-Gaussian Distributions}
In {several} experiments, particularly involving {nitrogen vacancy (NV)} centers, {it has been reported}
that the spin frequencies follow a $q$-Gaussian distribution~\cite{Putz2014,Krimer2014}, 
which is a general distribution function with a continuous parameter $q$ that connects it to a broad range of distributions. 
Some interesting cases {that} 
are relevant for modeling inhomogeneous spin ensembles, include the Lorentzian
$(q=2)$, Gaussian
$(q=1)$, Wigner semicircle
$(q=0)$, and two-point distribution function $(q=-1)$ as shown in Fig.~\ref{fig:q-bounds_q_var}(a). Therefore, solving the $q$-Gaussian frequency distribution allows us to tackle a wide range of spin ensembles}. The distribution function for $-1\leq q \leq 1$ is given as per Askey scheme \cite{Koekoek2010}
\begin{align}
    P_q(x)&=\frac{\sqrt{1-q}}{2\pi}\sqrt{4-(1-q)x^2}\prod_{k=1}^{\infty}w_k(q),~\textrm{where}\label{eq:q-dist}\\
    w_k(q)&=\left[1-(1-q)x^2(1-q^k)q^{k-1}+(1-q^2)q^{2k-1}\right]^{-1}\nonumber
\end{align}
{and} $x=(\omega-\bar{\omega})/\sigma_\omega$.
{There are two regimes of interest: i) 
$-1\leq q\leq 1$, 
which is a bounded probability distribution 
with 
a cutoff frequency, and ii) $1<q<2$, which represent heavy tail distributions ranging from 
Gaussian to Lorentzian distributions. 
The two regimes can be solved separately.}

\subsubsection{Finite frequency range $(-1\leq q\leq 1)$}
The corresponding orthogonal set of polynomials associated with {$P_q (x)$ in this range are} $q$-Hermite polynomials
\begin{equation}
\sum_{n=0}^{\infty} \frac{H_n(x|q)}{(q;q)_n} t^n
= \frac{1}{(t e^{i\theta}, t e^{-i\theta}; q)_\infty},
\end{equation}
{where $x=\cos\theta$ and}
$(x;q)_n=\prod_{k=0}^{n-1}(1-x q^k)$ is the $q$-Pochhammer symbol and $(x,y;q)_n=(x;q)_n (y;q)_n$. This gives the first two polynomials as
\begin{align}
H_0(x|q) &= 1~~\textrm{and}~~ H_1(x|q) = 2x
% ...
\end{align}
and for $n \geq 1$, a recurrence relation of the form
\begin{equation}
H_{n+1}(x|q) = 2x H_n(x|q) - (1-q^{n}) H_{n-1}(x|q).
\label{recursion-qGauss}
\end{equation}
{An orthonormal polynomial set $\{\pi_n(x)\}$ can be constructed by limiting the range of integration for variable $x=\omega-\bar{\omega}$ to $\{-x_0,x_0\}$, where $x_0=2\sigma_\omega/\sqrt{1-q}$, such that}
\begin{align}
&\pi_n(x)=\frac{1}{\sqrt{g_{\rm{eff}}\mathcal{N}_q}}H_{n}\left(\frac{x\sqrt{1-q}}{2\sigma_\omega}|q\right),~\textrm{where}\\
&\int_{-x_0}^{x_0}\pi_n(x)\pi_m(x)P_q(x)dx=\delta_{n,m}.
\end{align}
{Therefore, using the recursion relation in Eq.~\eqref{recursion-qGauss}, the relevant terms can be derived as}
\begin{align}
\beta_n=\sigma_\omega\sqrt{n_q},~\textrm{where}~n_q=\frac{1-q^n}{1-q},
\label{beta-qGauss}
\end{align}
{and $\alpha_n=\bar{\omega}$, {for} $n\geq1$. Using the above 
terms, the 
Hamiltonian $\mathcal{H}_k$ can be analytically derived in the Krylov space.
In the limit $q\rightarrow1^{-}$, where $n_q\rightarrow n$, the distribution reduces to a Gaussian distribution.
While $\beta_n$ increases with $n$ for Gaussian distribution, for large $n$ and $q<1$, $\beta_n\rightarrow\sigma_\omega/\sqrt{1-q}$.
Therefore, the Lieb-Robinson bound is finite, and is given by 
$v_{\rm{max}}=2\sigma_\omega/\sqrt{1-q}$.}

{Figure~\ref{fig:q-bounds_q_var} shows the spread of a single excitation, initially stored in the bright state $\ket{\phi_1}$, through the Krylov space of a spin ensemble with a $q$-Gaussian distribution of spin frequencies.
For $q=1$, the distribution is Gaussian, and strong parabolic dispersion of information to higher Krylov states is observed, as shown in Fig.~\ref{fig:q-bounds_q_var}(b). Importantly, there is no revival in population of the bright state $\vert \phi_1\rangle$, 
as the coupling term $\beta_{n+1}>\beta_{n}$ increases with $n$, leading to unbounded Lieb-Robinson velocities and flow of information away from the initial bright state. 
In contrast, for $q=0$ in Fig.~\ref{fig:q-bounds_q_var}(c), the information spreads linearly, with finite revival in bright mode population. This is evident by the fact that the coupling $\beta_n = \sigma_\omega$ is independent of $n$ and the probability of information transfer to a higher Krylov state is the same as flow to a lower state.  
The Lieb-Robinson bound is given by $v_{\rm{max}}=2\sigma_\omega$.}
The revivals are much stronger
{for} $q=-0.5$, as shown in Fig.~\ref{fig:q-bounds_q_var}(e), and eventually becomes
close to Rabi-like oscillations
as $q\rightarrow-1$ in Fig.~\ref{fig:q-bounds_q_var}(f). 
{From Eq.~\eqref{beta-qGauss}, it is evident that $\beta_n = 0$ and $\sigma_\omega$, for even and odd Krylov states, respectively, when $q\rightarrow -1$. As such, the coupling between $\ket{\Phi_2}$ and $\ket{\Phi_3}$ vanishes, thus preventing information leakage to higher Krylov states.
The values of $\beta_n$ for different $q$ are also highlighted in Fig.~\ref{fig:q-bounds_q_var}(d).}
The finite revivals in the free evolution of bright mode are important as they imply localized dynamics, and one can retrieve the lost information back using optimal control protocols. 
%
{The dynamics of a spin ensemble with a $q$-Gaussian distribution of spin frequencies can be analytically mapped to $q$-oscillator algebra, as shown in Appendix~\ref{app:q-oscillator}, which allows for an elegant formalism to study the flow of information.}

\subsubsection{Infinite frequency range $(1<q<3)$}

In this limit, {the distribution function is often represented by the Tsallis q-Gaussian function \cite{Umarov2008}}

\begin{align}
    P_q(\xi)&=\frac{\sqrt{\beta}}{C_q}\left[1-(1-q)\beta \xi^2\right]^{1/(1-q)},\label{eq:Tsallis_qGauss}
\end{align}
where,
\begin{align}
    C_q&=\frac{\sqrt{\pi}\Gamma\left(\frac{3-q}{2(1-q)}\right)}{\sqrt{q-1}\Gamma\left(\frac{1}{1-q}\right)}.
\end{align}
The associated orthogonal polynomial set to this  
distribution is
the relativistic Hermite polynomials~\cite{Vignat2011-ll}, which 
is defined for $\xi=\sqrt{\beta}(\omega-\bar{\omega})$ as
\begin{align}
H^{(N)}_n(\xi)=\frac{(2N)_n}{(2\sqrt{N})^n}&\sum_{k=0}^{\lfloor n/2\rfloor}\left[\frac{(-1)^k}{(N+\frac{1}{2})_k}\times\right.\nonumber\\
&\left.\frac{n!}{(n-2k)!k!}\left(\frac{2\xi}{\sqrt{N}}\right)^{n-2k}\right].
\end{align}
Here, $()_k$ is the Pochhamar symbol and 
{$N=1/(q-1)$, for $1<q<3$.
For
$q\rightarrow1^+$, $N\rightarrow\infty$, 
which results in 
$\text{lim}_{\lambda\rightarrow\infty}H^{(N)}_n(\xi)\rightarrow H_n(\xi)$, and the Hermite polynomials for the Gaussian distribution is recovered.}  
{An orthonormal set $\{\pi_n^N(\xi)\}$ is then derived as}
\begin{align}
\pi^{(N)}_n(\xi)=\frac{(1+\xi^2/N)^{-(n+1)/2}}{\sqrt{\mathcal{N}_n}}H^{(N)}_n(\xi),
\end{align}
where
\begin{align}
&\int_{-\infty}^\infty \pi^{(N)}_n(\xi)\pi^{(N)}_m(\xi)P_q(\xi)dx=\delta_{m,n},~\textrm{and}\\
&\mathcal{N}_n=\frac{\sqrt{N}2^n n!~\Gamma(N)\Gamma(N-1/2)}{\Gamma(N+n)}.
\end{align}

{In this regime, to estimate the terms $\alpha_n$ and $\beta_n$ in the Krylov Hamiltonian, we make use of the moments of the distribution~\cite{MAGNUS2008} and the Henkel determinant approach in Appendix~\ref{app:henkel}.
It is known that these moments remain finite only for $n<N-1/2$~\cite{Tsallis1995,MAGNUS2008}. This is due to the fact that the distributions are heavy-tailed for $q >1$, which makes an effective Krylov basis representation difficult in this regime.}


\subsection{Uniform Distribution}

In 
{several spin ensembles, the distribution of spin frequencies} can be well approximated
{by} a uniform distribution over a finite range of frequencies. 
A uniform frequency distribution is given by}
\begin{align}
    P(\omega)&=\begin{cases}
                \frac{1}{2\sqrt{3}\sigma_\omega},~~~~|\omega-\bar{\omega}|\leq\sqrt{3}\sigma_\omega\\
                0~~~~~~~,~~~~~\rm{otherwise}
            \end{cases},
\end{align}
where $\int_{-\infty}^{\infty}P(\omega)d\omega=1$. In this case, the orthogonal polynomial set $\{\pi_n(\omega)\}$ 
{is the} Legendre polynomials, defined as
\begin{align}
    \pi_n(\omega)&=\sqrt{2n+1}~P_n\left(\eta\right),~n=0,1,2,3,...,
\end{align}
where $\eta=(\omega-\bar{\omega})/\sqrt{3}\sigma_\omega$. 
%
%
The Legendre polynomials follow a recursion relation
\begin{align}
    (n+1)P_{n+1}(\eta)=(2n+1)\eta P_n(\eta)-nP_{n-1}(\eta).
\end{align}
Thus, re-expressing it in terms of $\{\pi_n(\omega)\}$:
\begin{align}
    (\omega-\alpha_n)\pi_n(\omega)&=\beta_{n+1}\pi_{n+1}(\omega)+\beta_n\pi_{n-1}(\omega),\\
    \alpha_n=\bar{\omega}&,~~\beta_n=\frac{\sqrt{3}\sigma_\omega n}{\sqrt{(2n+1)(2n-1)}}.
    \label{beta-uniform}
\end{align}
{From Eq.~\eqref{beta-uniform} above, it is evident that $\beta_1 = \sigma_\omega$ and $\beta_n\rightarrow\sqrt{3}\sigma_\omega/2$ for large $n$. As such, unlike Gaussian distribution, the coupling converges at high $n$, with a finite Lieb Robinson bound given by $v_{\rm{max}}=2\sigma_\omega$. This implies a linear bound on the spread on information as shown in Fig.~\ref{fig:uniform}.}



\begin{figure}[t]
    \centering
    \includegraphics[width=\linewidth]{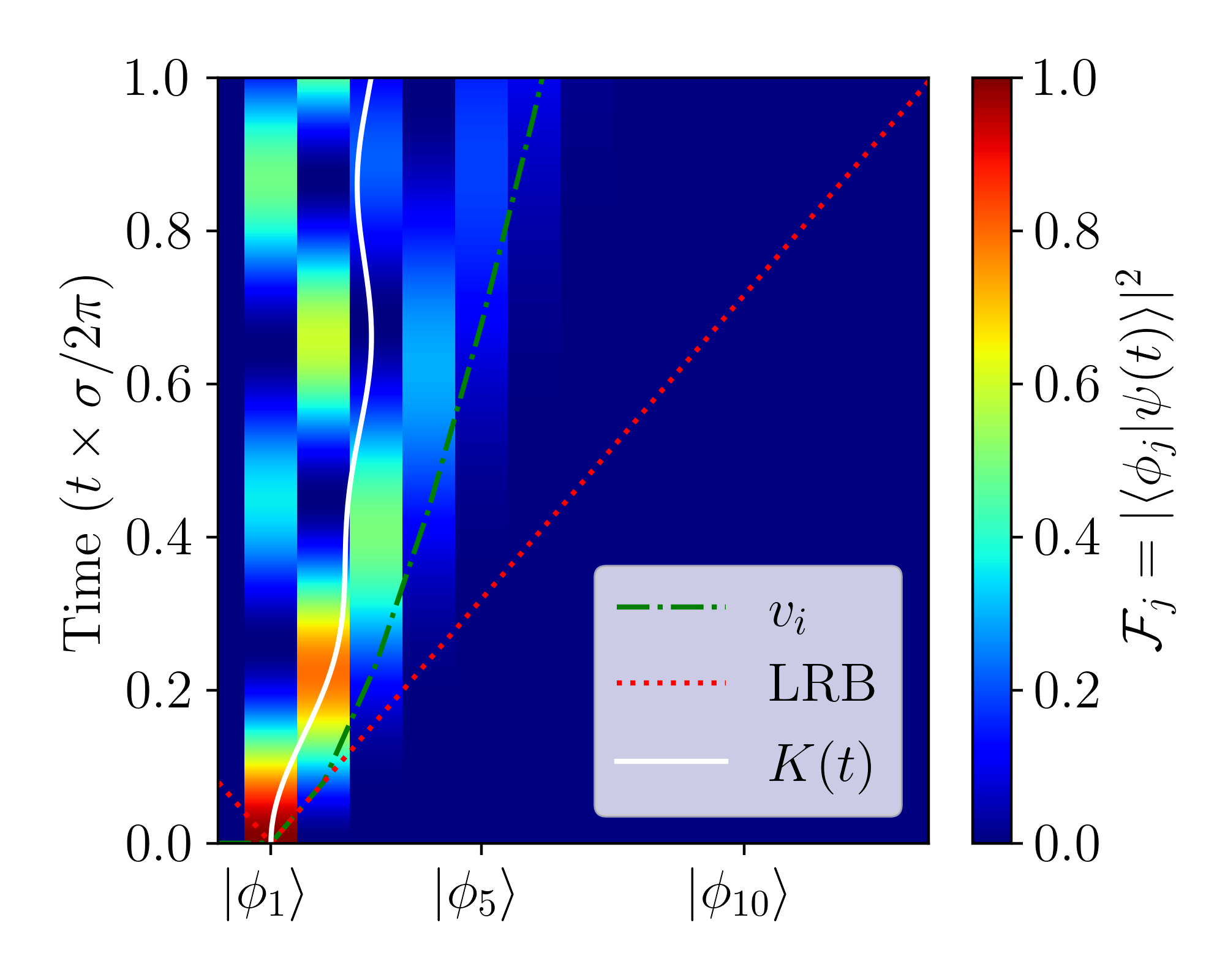}
    \caption{Fidelity evolution from bright mode $\vert\phi_1\rangle$ in a spin ensemble with uniform spin frequency distribution.}
    \label{fig:uniform}
\end{figure}

\section{Mapping to Physical Systems \label{sec:physical}}

{Spin ensembles are an integral component of hybrid quantum computing architectures~\cite{Kurizki2015}, and platforms exhibit a wide range of distribution of spin frequencies, ranging from naturally occurring to engineered spins. The most widely occurring spin ensembles are modeled using the Gaussian distribution, including those used in studying nuclear magnetic and electron spin
resonance~\cite{Stanley2014}} and thermal Doppler broadening in atomic ensembles~\cite{Lugiato2015}. The Krylov formalism was used to design optimally controlled, long time storage of a qubit in a Gaussian spin ensemble~\cite{Gupta2026a}.
%

{The related family of $q$-Gaussian spin frequency distributions has been used for modeling spin frequency inhomogeneity in NV centers~\cite{Putz2014,Krimer2019}. Interestingly, in the context of engineered spin ensembles such as spectral hole burning~\cite{Putz2017,El-Ella2019}, the effective spin frequency distribution can be well approximated by a $q$-Gaussian distribution with $q<0$ (see Fig.~\ref{fig:q-bounds_q_var}(a)) and provides a single-parameter mapping of Gaussian distributions to a two-peak spin frequency distribution. In NV centers, $q$-Gaussian distributions for $q>1$ have also been studied~\cite{Sandner2012, Krimer2019}. The $q=0$ case maps to the Wigner semicircle distribution, which arises in the case of highly interacting thermal systems where the system behavior can be effectively mapped to random matrix theory~\cite{wigner1967,Erds2008}. 
}

{Spin ensembles with uniform distribution often arise in systems that are spectrally engineered~\cite{Rubio2018,Debnath2019} for implementing sensing or quantum memory protocols. These include rare-earth-ion doped solid state spins~\cite{Laplane2015} and NV centers in diamond~\cite{Putz2017}. Other models with uniform distribution include atomic frequency combs, where multiple teeth of the comb can represent a different uniform frequency distribution~\cite{Dhar2018, Zens2021}.}

\section{Conclusion and Outlook\label{sec:conc}}
In this work, {a unified theoretical} framework {is presented} to study {the spread of} quantum information 
in 
inhomogeneous spin ensembles.
By constructing the exact Krylov {basis in the single-excitation, fully quantum regime, the Hamiltonian and the subsequent dynamics}
of an inhomogeneous {spin ensemble is mapped to}
an effective one-dimensional Krylov lattice. 
{This enables an exact description of information flow and spread in the effective Hilbert space using only the statistical properties of the spin ensemble. This allows for derivation of analytical expressions for the Lieb-Robinson velocity and the Krylov complexity, simply using the terms of the Hamiltonian in the Krylov basis. Moreover, the optimal time of state transfer between the Krylov states, governed by Mandelstam–Tamm bound for the quantum speed limit, can also be estimated. 
%
%
As a result, a detailed, state-resolved characterization of quantum information flow in the single-excitation space is possible, well beyond mean-field descriptions~\cite{Sandner2012,Sharma2025}, for inhomogeneous ensembles with  different distributions of spin frequencies.}

{Two key outcomes from the study is firstly, the direct dependence of the flow of information in the Krylov subspace to the derived off-diagonal or coupling terms of the Hamiltonian, which is critical for optimal control~\cite{Gupta2026a}. Secondly, these coefficients can be calculated from the statistical properties of the spin distribution viz. the mean, standard deviation or the moments. It is shown that for Gaussian distributions, the coupling terms increase for higher Krylov states, which leads to rapid dispersion of information to higher Krylov states. On the other hand, for $q$-Gaussian with $q=0$ and uniform distributions, the terms are constant at large $n$, leading to population revivals and information backflow. {The spread and localization of information are related to $q > 0$ and $q < 0$, respectively, with stronger behavior away from $q=0$. 
For example, engineered ensembles with $q\rightarrow -1$, can have a dramatic effect, with a complete cutoff of information flow to higher Krylov states, resulting in very strong localization.}

{The framework} opens up several promising directions. The exact Krylov-space construction presented can be extended to {higher excitation subspaces},
where spin-interaction and genuine many-body effects may become important. It also provides a natural starting point for studying the interplay between inhomogeneity, dissipation, and measurement, which are highly relevant 
{implementation of quantum protocols using spin ensembles.}
Moreover, the connection between distribution geometry, $q$-deformation, and information localization suggests new strategies for engineering spin ensembles with tailored spectral properties to optimize quantum memory and information protection protocols. 
More broadly, our 
The study 
{reiterates the strength} of Krylov-space based methods as a powerful analytical tool for understanding quantum dynamics in disordered and structured {quantum, solid-state platforms.}

\begin{acknowledgments}
R.G. acknowledges funding from CSIR-HRDG, India in the form of Senior Research Fellowship and thanks Imperial College London for support during a visit through the Global Development Hub Fellows Fund.
H.S.D. acknowledges financial support from SERB-DST, India via a
Core Research Grant (No: CRG/2021/008918) and the Industrial Research \& Consultancy Centre, IIT Bombay via grant RD/0521-IRCCSH0-001 (No: 2021289).
\end{acknowledgments}

\onecolumngrid
\appendix

\section{Obtaining recursion coefficients from moments\label{app:henkel}}

In Sec.~\ref{sec:Krylov_coeff}, {it is shown that the terms in the Krylov Hamiltonian $\alpha_n$ and $\beta_n$}  
can be calculated using the orthogonal polynomials associated with the
{probability distribution of the spin frequency.}
However, in cases where the distribution function is not in a standard function and the exact form of orthogonal polynomials are not known, {but the moments of the distribution can be readily calculated, then the}
Krylov terms can be calculated using the 
Hankel determinant approach~\cite{Liu2021}. 
{If $\{m_k=\textbf{E}(\Delta^k)\}_{k\ge0}$ are the finite moments of the distribution}, the
Hankel matrix can be defined as
\begin{align}
{h_n} :&=
\begin{pmatrix}
m_0 & m_1 & \cdots & m_{n-1}\\
m_1 & m_2 & \cdots & m_n\\
\vdots & \vdots & \ddots & \vdots\\
m_{n-1} & m_n & \cdots & m_{2n-2}
\end{pmatrix}, ~\textrm{where}~~
D_{-1}=1, D_0=1, D_n:=\det ({h_n}), n\in\mathbb{N}.
\end{align}
%
{Note that $m_n=0~\forall~n\in\rm{odd}$ holds for all centrally, symmetric spin distribution.}
To extract the recursion relation of the coefficients $\alpha_n$ and $\beta_n$, as presented in Eqs.~\eqref{eq:three_term_rec}-\eqref{eq:hankel_formulae}, the $(n+1)^{\rm{th}}$ terms are obtained by applying
Cramer's rule~\cite{Arfken2013}. The Hankel matrix
$h_n$ is modified by replacing the last column with the
last column of $h_{n+1}$ (with the trailing entry removed) 
and denoting the resulting determinant as
\begin{equation}
D'_n=
\begin{vmatrix}
m_0 & m_1 & \cdots & m_{n-2} & m_n\\
m_1 & m_2 & \cdots & m_{n-1} & m_{n+1}\\
\vdots & \vdots &        & \vdots & \vdots\\
m_{n-1} & m_n & \cdots & m_{2n-3} & m_{2n-1}\end{vmatrix},\quad n\ge2,
\end{equation}
where $D'_0=0$ and $D'_1=m_1$.
{Now, the terms $\alpha_n$ and $\beta_n$,}
which are the diagonal and off-diagonal elements of the {Hamiltonian in the Krylov space}
{can be} obtained as
\begin{align}
\alpha_n&=\frac{D'_n}{D_n}-\frac{D'_{n-1}}{D_{n-1}},~
\beta_0=\sqrt{m_0}=\sqrt{\sum_j g_j^2},\nonumber\\
\beta_n&=\sqrt{\frac{D_{n+1}D_{n-1}}{D_n^2}},~ n\in\mathbb{N}.
\label{eq:HenkelCoeff}
\end{align}

{The Hankel approach can be applied to the Gaussian spin distribution, where $\chi_k$ {in Eqs.~\eqref{chi_1}-\eqref{chi_2}} is equal to $\sigma_\omega^{2k} k!$~\cite{gautschi1996}. This gives}
\begin{align}
    D_n&=\prod_{k=0}^{n-1}\chi_k=\prod_{k=0}^{n-1}\sigma_\omega^{2k}k!=\sigma_\omega^{n(n-1)}\prod_{k=0}k!,\\
    \beta_n&=\sigma_\omega\sqrt{\frac{(D_n n!)(D_n/(n-1)!)}{D_n^2}}=\sigma_\omega\sqrt{n},~\forall~n\geq1.
\end{align}
{Here} $D_{n+1}=D_n n!$ and $D_{n-1}=D_n/(n-1)!$. The other elements are  $\alpha_0=\omega_c$ and $\alpha_n=\bar{\omega}$, {and $\beta_0$ is set as the collective coupling $g_{\rm{eff}}$. This leads to $\beta_n=\sigma_\omega\sqrt{n}$.}

{The second example is}
the $q$-Gaussian distribution defined in \eqref{eq:Tsallis_qGauss} with $q>1$. 
This is again a centrally symmetric distribution, where the odd moments vanish. 
However, the even moments can be derived, {as shown in Ref.~\cite{MAGNUS2008}}, by the relation 
\begin{align}
\mu_{2n}&=\int_{-\infty}^{\infty}C_q \xi^{2n}(1+\xi^2/N)^{-N}d\xi,\nonumber\\
\mu_{2n}&=C_q\sqrt{N}^{2n+1}\frac{\Gamma(N-(n+1)/2)\Gamma(n+1/2)}{\Gamma(N)}.
\end{align}
Using this, {the first few terms in the Hamiltonian are given by}
\begin{align}
\sigma_\omega=\beta_1=\frac{\mu_2}{\mu_0}=\frac{\beta N\Gamma(N-1)}{2\Gamma(N-1/2)},~\beta_2=\frac{\mu_4}{\mu_2}-\frac{\mu_2}{\mu_0}.
\end{align}
{However,} for $n\geq N-1/2$ the moments diverge, leading to truncation of the series, thus rendering the Krylov space formalism to remain valid only up to first few states. For general distribution, these coefficients can be quickly computed numerically using the Stieltjes procedure \cite{gautschi1996}.

\section{Correspondence to $q$-oscillator and exact unitary evolution\label{app:q-oscillator}}
{The Krylov space Hamiltonian for a spin ensemble with a $q$-Gaussian distribution of spin frequencies can be mapped directly to a $q$-analog of the quantum harmonic oscillator, defined by}
\begin{align}
    H_q=\sigma_\omega\sum_{n=0}^{M-1} \sqrt{n_q}X_{n,n+1}=\sigma_\omega\left(a_q + a^\dag_q\right)=\sigma_\omega x_q.
\end{align}
{Here, the $q$-boson annihilation and creation operators, $a_q$ and $a_q^\dag$ are defined as}
\begin{align}
a_q\vert\phi_n\rangle=\sqrt{n_q}\vert\phi_{n-1}\rangle~\textrm{and}~
a^\dag_q\vert\phi_n\rangle&=\sqrt{n_q + 1}\vert\phi_{n+1}\rangle,
\end{align}
where $a_q a^\dag_q - q a^\dag_q a_q=1$.
The propagator can thus be calculated using the spectral theorem as an integral over the $q$-Hermite polynomial set
\begin{align}
    U_q(t)&=e^{-iH_q t}=e^{-i\sigma_\omega(a_q + a^\dag_q)t},\\
    \left[U_q(t)\right]_{n,m}&=\int_{-x_0}^{x_0}e^{-ix_q t}\pi_n(x)\pi_m(x)P_q(x)dx.
\end{align}
These expressions even further simplify for $q=1$ (Gaussian distribution) to give a closed form of the propagator representation using the Baker-Campbell-Hausdorff relation
\begin{align}
    U_{q=1}(t)=e^{-\frac{1}{2}\sigma_\omega^2 t^2}e^{-i\sigma_\omega t a^\dag}e^{-i\sigma_\omega t a}.
\end{align}
This clearly explains the initial close-to-Gaussian decay of any excitation in the spin ensemble, which has been seen in most experiments. The matrix elements of this operator can be obtained by mapping to the $q$-oscillator, and for $q=1$, the usual Harmonic oscillator operator algebra can be utilized. The required matrix elements are
\begin{equation}
    U_{nm}(t)=\bra{\phi_n}U_{q=1}(t)\ket{\phi_m}=e^{-\frac{1}{2}\sigma_\omega^2 t^2}\bra{\phi_n}e^{-i\sigma_\omega t a^\dag}e^{-i\sigma_\omega t a}\ket{\phi_m}
\label{eq:unitary_comp}
\end{equation}
The annihilation and creation operators are operated using
\begin{align}
    e^{-i\sigma_\omega t a}\ket{\phi_m}=\sum_{j=0}^{\infty}\frac{(-i\sigma_\omega t)^{j}}{j!}a^j\ket{\phi_m}=\sum_{j=0}^{\infty}\frac{(-i\sigma_\omega t)^{j}}{j!}\sqrt{\frac{m!}{(m-j)!}}\ket{\phi_{m-j}}.
\label{eq:ket}
\end{align}
Similarly,
\begin{align}
    \bra{\phi_n}e^{-i\sigma_\omega t a^\dag}=\sum_{k=0}^{\infty}\bra{\phi_n}\left(a^\dag\right)^k\frac{(-i\sigma_\omega t)^{k}}{k!}=\sum_{k=0}^{\infty}\frac{(-i\sigma_\omega t)^{k}}{k!}\sqrt{\frac{n!}{(n-k)!}}\bra{\phi_{n-k}}.
\label{eq:bra}
\end{align}
The non-zero contribution in Eq.~\eqref{eq:unitary_comp} comes only when $m-j=n-k$. Defining $r=m-j$ and substituting Eq.~\eqref{eq:ket} and Eq.~\eqref{eq:ket} in Eq.~\eqref{eq:unitary_comp}, and assuming $n\geq m$,
\begin{align}
    U_{n,m}(t)=e^{-\frac{1}{2}\sigma_\omega^2 t^2}\sqrt{\frac{m!}{n!}}(-i\sigma_\omega t)^{n-m}\sum_{r=0}^m {^n}C_{m-r}\frac{\left(-\sigma_\omega^2t^2\right)^r}{r!},~~{^n}C_{m-r}=\frac{n!}{(m-r)!(n-m+r)!}.
\label{eq:raw_unitary}
\end{align}
Now, the expansion of generalized Laguerre Polynomials (Sec. 22.3.9 of~\cite{Abramowitz1964}) is used to express
\begin{align}
    L^{(n-m)}_m\left(\sigma_\omega^2 t^2\right)=\sum_{r=0}^m {^n}C_{m-r}\frac{\left(-\sigma_\omega^2t^2\right)^r}{r!},
\label{eq:Laguerre_Id}
\end{align}
and substitution of Eq.~\eqref{eq:Laguerre_Id} in Eq.~\eqref{eq:raw_unitary} is done to get a compact form
\begin{align}
    U_{n,m}(t)=e^{-\frac{1}{2}\sigma_\omega^2 t^2}\sqrt{\frac{m!}{n!}}(-i\sigma_\omega t)^{n-m}L^{(n-m)}_m\left(\sigma_\omega^2 t^2\right),
\label{eq:n_geq_m_unitary}
\end{align}
Similarly for $m\geq n$, one can also obtain a similar expression
\begin{align}
    U_{n,m}(t)=e^{-\frac{1}{2}\sigma_\omega^2 t^2}\sqrt{\frac{n!}{m!}}(-i\sigma_\omega t)^{m-n}L^{(m-n)}_n\left(\sigma_\omega^2 t^2\right),
\label{eq:m_geq_n_unitary}
\end{align}
which is expected as $U_{n,m}=U_{m,n}$, because the Krylov space Hamiltonian itself is a symmetric matrix.
Thus the combining Eq.~\eqref{eq:n_geq_m_unitary} and Eq.~\eqref{eq:m_geq_n_unitary}, a general expression is obtained as
\begin{align}
    U_{nm}(t)&=e^{-\frac{1}{2}\sigma_\omega^2 t^2}r_{n,m}(-i\sigma_\omega t)^{\vert n-m\vert}L^{(\vert n-m\vert)}_{\rm{min}(n,m)}\left(\sigma_\omega^2t^2\right),
    \label{eq:unitary_gaussian}
    \\ &\textrm{where}~~
    r_{n,m}=\sqrt{\frac{\rm{min}(n,m)!}{\rm{max}(n,m)!}},\nonumber\label{eq:final_unitary}
\end{align}
This formulae gives the time evolution analytically, without numerically solving the Schr\"{o}dinger equation.
The expression allows the exact evolution coefficients from any initial state to any
time $t$. The exact expression for Krylov complexity can be determined this way using Eq.~\eqref{eq:krylov_complexity}. For initialization with the bright mode, $\vert\psi(0)\rangle=\vert\phi_1\rangle,~c_n(0)=\delta_{1,n}$, $c_n(t)$ is obtained as
\begin{align}
    c_n(t)&=\sum_m U_{n,m}(t)c_m(0)=\sum_m U_{n,m}(t)\delta_{1,m}=U_{n,1}(t),\\
    c_n(t)&=e^{-\frac{1}{2}\sigma_\omega^2t^2}r_{n,1}(-i\sigma_\omega t)^{|n-1|}L^{(|n-1|)}_{\rm{min(n,1)}}(\sigma_\omega^2t^2).
\end{align}
Thus, population of Bright mode at any given Krylov state $\ket{\phi_n}$ can be easily computed as $|\langle\phi_n\vert\psi(t)\rangle|^2=|c_n(t)|^2$.



\twocolumngrid
\bibliography{Bibliography_file}

\end{document}